\documentclass[pre,,noshowpacs,showkeywords,twocolumn, amsmath, amssymb, nodata,nofootinbib]{revtex4}
\usepackage{graphics}
\usepackage{epsfig}
\usepackage{graphicx}
\usepackage{dcolumn}
\usepackage{bm}

\usepackage{amsmath}
\usepackage[english]{babel}
\usepackage{color}

\begin{document}

\title{Elastic Properties and Glass Forming Ability of the \\ Zr$_{50}$Cu$_{40}$Ag$_{10}$ Metallic Alloy}

\author{Ramil~M.~Khusnutdinoff$^{1,2 a}$ and Anatolii~V.~Mokshin$^{1,2 b}$}

\affiliation{$^1$Kazan Federal University,
 \it 420008 Kazan, Russia,\\
   $^2$Udmurt Federal Research Center of the Ural Branch of the Russian Academy of Sciences, 
   \it  426068 Izhevsk, Russia \\
   {\small \rm E-mail: $^{a}$khrm@mail.ru \;  $^{b}$anatolii.mokshin@mail.ru}}

\begin{abstract}
The elastic properties of the Zr$_{50}$Cu$_{40}$Ag$_{10}$ metallic alloy, such as the bulk modulus $B$, the shear modulus $G$, the Young's modulus $E$ and the Poisson's ratio $\sigma$, are investigated by molecular dynamics simulation in the temperature range $T=250-2000$~K and at an external pressure of $p = 1.0$ bar. It is shown that the liquid-glass transition is accompanied by a considerable increase in the shear modulus $G$ and the Young's modulus $E$ (by more than $50\%$). The temperature dependence of the Poisson's ratio exhibits a sharp fall from typical values for metals of approximately $0.32-0.33$ to low values (close to zero), which are characteristic for brittle bulk metallic glasses. Non-monotonic temperature dependence of the longitudinal and transverse sound velocity near the liquid-glass transition is also observed. The glass forming ability of the alloy is evaluated in terms of the fragility index $m$. As found, its value is $m\approx64$ for the Zr$_{50}$Cu$_{40}$Ag$_{10}$ metallic glass, that is in a good agreement with the experimental data for the Zr-Cu-based metallic glasses.
\end{abstract}

\keywords{molecular dynamics, elastic moduli, amorphous metallic alloy, structural transformations, glass forming ability}


\maketitle

\section{Introduction}

Interest to the amorphous metallic alloys is provoked by the obvious advantages of their potential applications in various fields of techniques, engineering and materials science. Amorphous alloys or metallic glasses are a new class of materials with unique physical and mechanical properties, which differ from the properties of their crystalline analogs~\cite{tempbib1, tempbib2}. So, for example, non-crystalline metallic alloys are characterized by high strength and elasticity, as well as good ductility under strong deformation effects. As a rule, metallic glasses are multi-component systems with high glass forming ability, the disordered phase of which can be formed by means of cooling the equilibrium melt with the cooling rates of $\gamma>10^{4}$~K/s~\cite{tempbib3}. Despite a significant number of experimental and numerical studies, there is still no clear understanding of the mechanisms of formation of structural heterogeneities and their influence on the elastic and mechanical properties of metallic glasses.

The aim of this work was a thorough study of the elastic properties of the Zr$_{50}$Cu$_{40}$Ag$_{10}$-alloy in a wide temperature range, which includes ``liquid-glass'' transition.

\section{Simulation Details}
Molecular dynamics (MD) simulations of the Zr$_{50}$Cu$_{40}$Ag$_{10}$ melt were carried out with an isothermal-isobaric ($NPT-$) ensemble in the temperature range $T=250-2000$~K at an external pressure of $p=1.0$~bar. To stabilize the temperature and the pressure of the system, the Nos\'{e}-Hoover thermostat with a relaxation parameter of $0.1$~ps and the barostat with damping parameter of $1.0$~ps were applied. The system was consisted of $N=32000$ atoms enclosed in a cubic cell with the periodic boundary conditions. The interactions between atoms are described by means of the EAM model potential~\cite{tempbib4}. Initial configurations were equilibrated Zr$_{50}$Cu$_{40}$Ag$_{10}$ melt at $T=2000$~K. Then, the system was cooled down to $T=250$~K with the cooling rate $1.0$~K/ps~\cite{tempbib5}. The integration of the equations of motion of atoms was carried out using the velocity Verlet algorithm with a time step of $\tau=1.0$~fs. To bring the system to a state of thermodynamic equilibrium, we performed $10^{6}$ time steps and then $10^{7}$ steps to calculate the mechanical properties and characteristics. All simulations were performed in the LAMMPS simulation package~\cite{tempbib6}.

\section{Results and Discussions}

\begin{figure}[ht]
	\centering
	\includegraphics[width=1.0\linewidth]{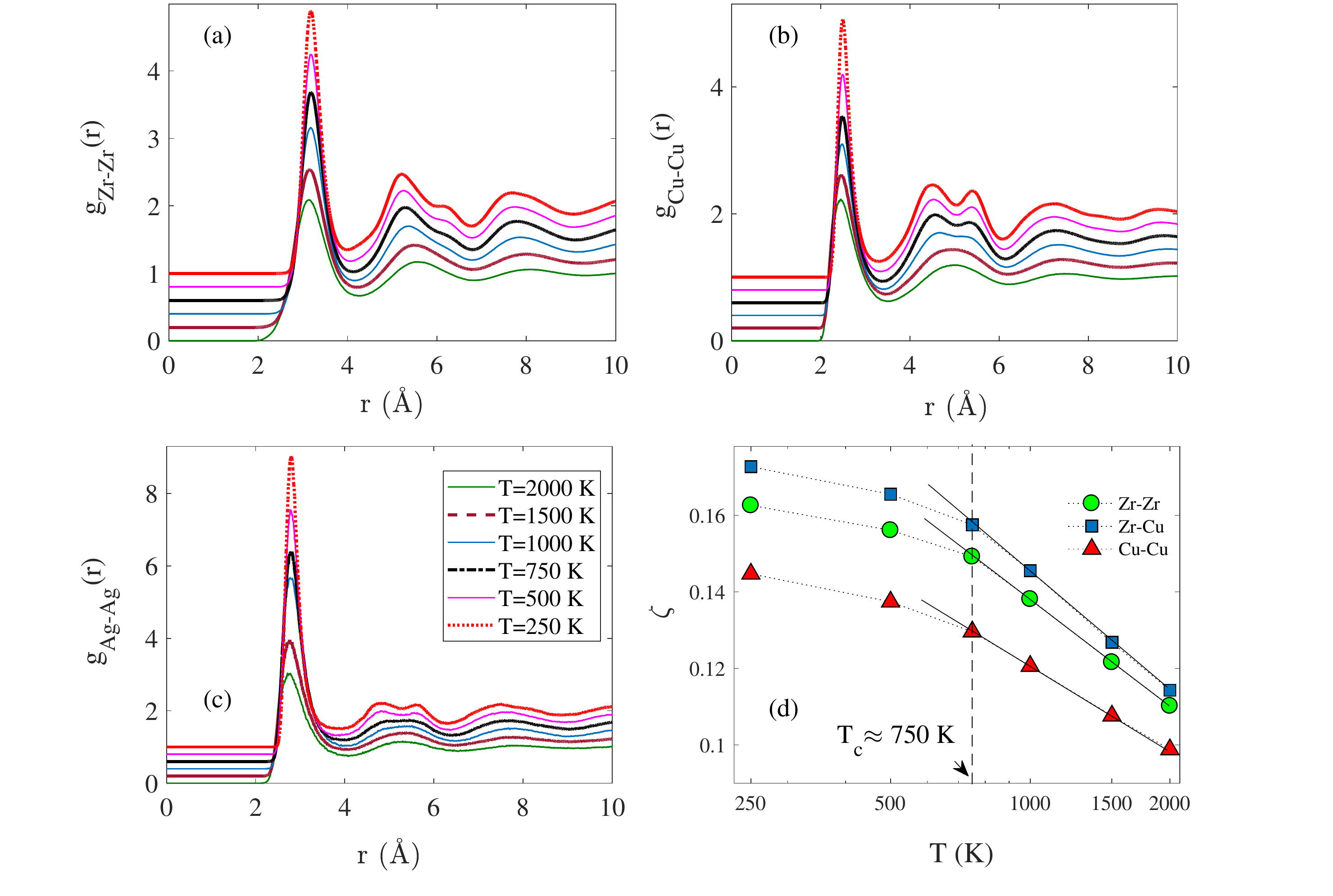}
	\caption{Partial components of the radial distribution function and the translational order parameter for Zr$_{50}$Cu$_{40}$Ag$_{10}$ melt at different temperatures.}
	\label{fig_1}
\end{figure}

Fig. $1$ (a-c) shows the partial components of the radial distribution function for the Zr$_{50}$Cu$_{40}$Ag$_{10}$ melt at various temperatures. As seen from the figure, all partial components $g_{\alpha,\beta}(r)$ (where $\alpha,\beta \in \{$Zr,~ Cu,~ Ag$\}$) with decreasing temperature manifest the structural transformations in the system due to the transition from the equilibrium melt to the amorphous glassy state~\cite{tempbib7}. The critical temperature $T_{c}$ associated with liquid-glass transition was determined using the translational order parameter~\cite{tempbib2}. The temperature dependence of this parameter in the semilogarithmic coordinates has two segments with an intersection point at $T_{c}\approx750$~ K (see Fig.~\ref{fig_1}d).

The elastic properties of the Zr$_{50}$Cu$_{40}$Ag$_{10}$ metal melt were studied by means of evaluation of the elastic (bulk and shear) moduli~\cite{tempbib8, tempbib9}. For the case of the $NPT$-ensemble, the isothermal bulk modulus is related to the mean-square fluctuations $\sigma_{V}^{\;2}$  of the volume $V$ of the simulation cell by the expression:
\begin{equation}
\label{eq:1.1}
      B = \frac{k_{B}TV}{\sigma_{V}^{\;2}}. \nonumber
\end{equation}

The shear modulus was calculated by the formula
\begin{equation}
\label{eq:1.2}
      G = \frac{V}{k_{B}T} \left\langle \left | p_{xy} (0) \right|^{2}  \right\rangle, \nonumber
\end{equation}
where angle brackets mean averaging over time samples and  $p_{xy} (0)$  are the non-diagonal components of the pressure tensor.

To assess the deformation properties of materials, it is necessary to know the equilibrium elastic modulus (the so-called Young's modulus) and the transverse strain coefficient (the so-called Poisson's ratio). The Young's modulus $E$ and the Poisson's ratio $\sigma$ are related with the elastic moduli as follows
\begin{equation}
\label{eq:1.3}
      E = \frac{9BG}{3 B + G}, \; \; \; \; \sigma = \frac{3B - 2G}{6B + 2G}. \nonumber
\end{equation}

\begin{figure}[ht]
	\centering
	\includegraphics[width=1.0\linewidth]{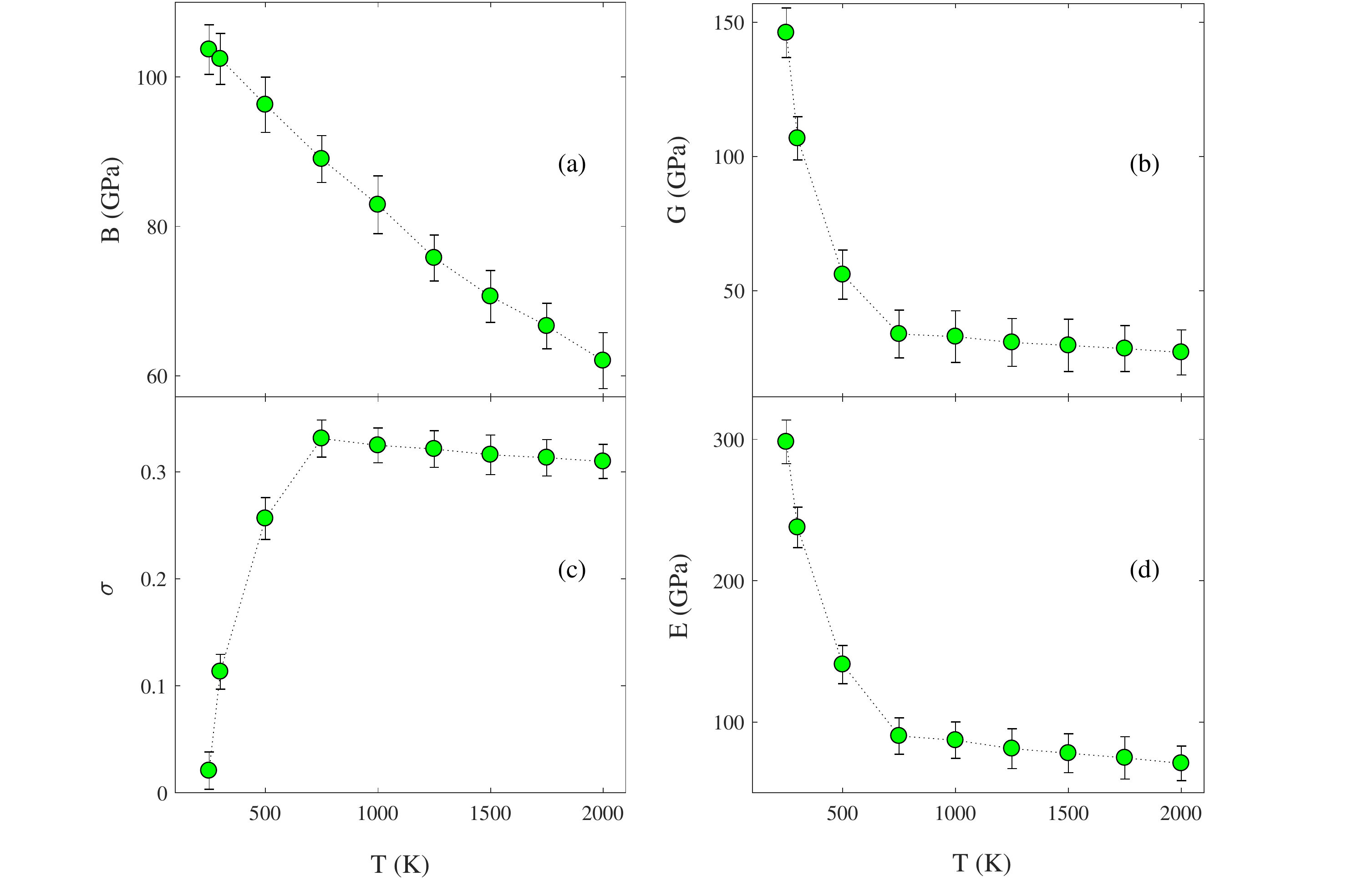}
	\caption{Temperature dependences of the bulk modulus $B$, the shear modulus $G$, the Poisson's ratio $\sigma$ and the Young's modulus $E$ for the Zr$_{50}$Cu$_{40}$Ag$_{10}$ metallic alloy. }
	\label{fig_2}
\end{figure}

The temperature dependences of the bulk modulus $B$, the shear modulus $G$, the Poisson's ratio $\sigma$ and the Young's modulus $E$ for the Zr$_{50}$Cu$_{40}$Ag$_{10}$ metallic alloy are plotted in Fig.~\ref{fig_2}. As seen from this figure, the temperature dependences of the shear modulus $G$, the Poisson's ratio $\sigma$ and the Young's modulus $E$ are characterized by a sharp change near the liquid-glass transition. At the same time, the bulk modulus grows monotonically with decreasing temperature. The importance of the Poisson's ratio in the study of glasses was pointed out by Novikov and Sokolov~\cite{tempbib10}. They found that the Poisson's ratio is related to the fragility index $m$ of the glass forming liquid, which can be used to account for the glass forming ability (GFA) properties of a melt. So, for example, the Poisson's ratio for brittle bulk metallic glasses demonstrates a low value. It can be seen from the Fig.~\ref{fig_2} that glassy Zr$_{50}$Cu$_{40}$Ag$_{10}$-alloy at the temperature $T=250$~K is characterized by a Poisson's ratio close to zero. The Poisson's ratio takes values close to zero showing very little lateral expansion when compressed. The fragility index $m$ was calculated by the formula~\cite{tempbib10}
\begin{equation}
\label{eq:1.4}
      m = 17 + 29 (B/G - 1), \nonumber
\end{equation}
where $17$ is the lowest value expected for $m$ and $B/G$ in glasses is not expected to be lower than $1$~\cite{tempbib11}. For the Zr$_{50}$Cu$_{40}$Ag$_{10}$-system, the calculated value of the fragility index is $m=64\pm4$, which is in a good agreement with the experimental data for the Zr-Cu-based metallic glasses (see Table~\ref{tab_KhRM}).

\begin{table}[h] 
\caption{Experimental data on the fragility index for amorphous Zr-Cu-based metallic alloys~\cite{tempbib12,tempbib13}.}
\begin{center} 
\begin{tabular}{ | c | c | }
\hline
Alloy composition & Fragility index, $m$  \\ \hline \hline
Zr$_{50}$Cu$_{50}$ & $76\pm10$ \\
\hline
Zr$_{44}$Cu$_{50}$Ti$_{6}$ & $48\pm4$ \\
\hline
Zr$_{46}$Cu$_{46}$Al$_{8}$ & $63\pm7$ \\
\hline
Zr$_{47}$Cu$_{47}$Al$_{6}$ &  $53\pm5$ \\
\hline
\end{tabular}
\end{center}
\label{tab_KhRM}
\end{table}

\begin{figure}[ht]
	\centering
	\includegraphics[width=1.5\linewidth]{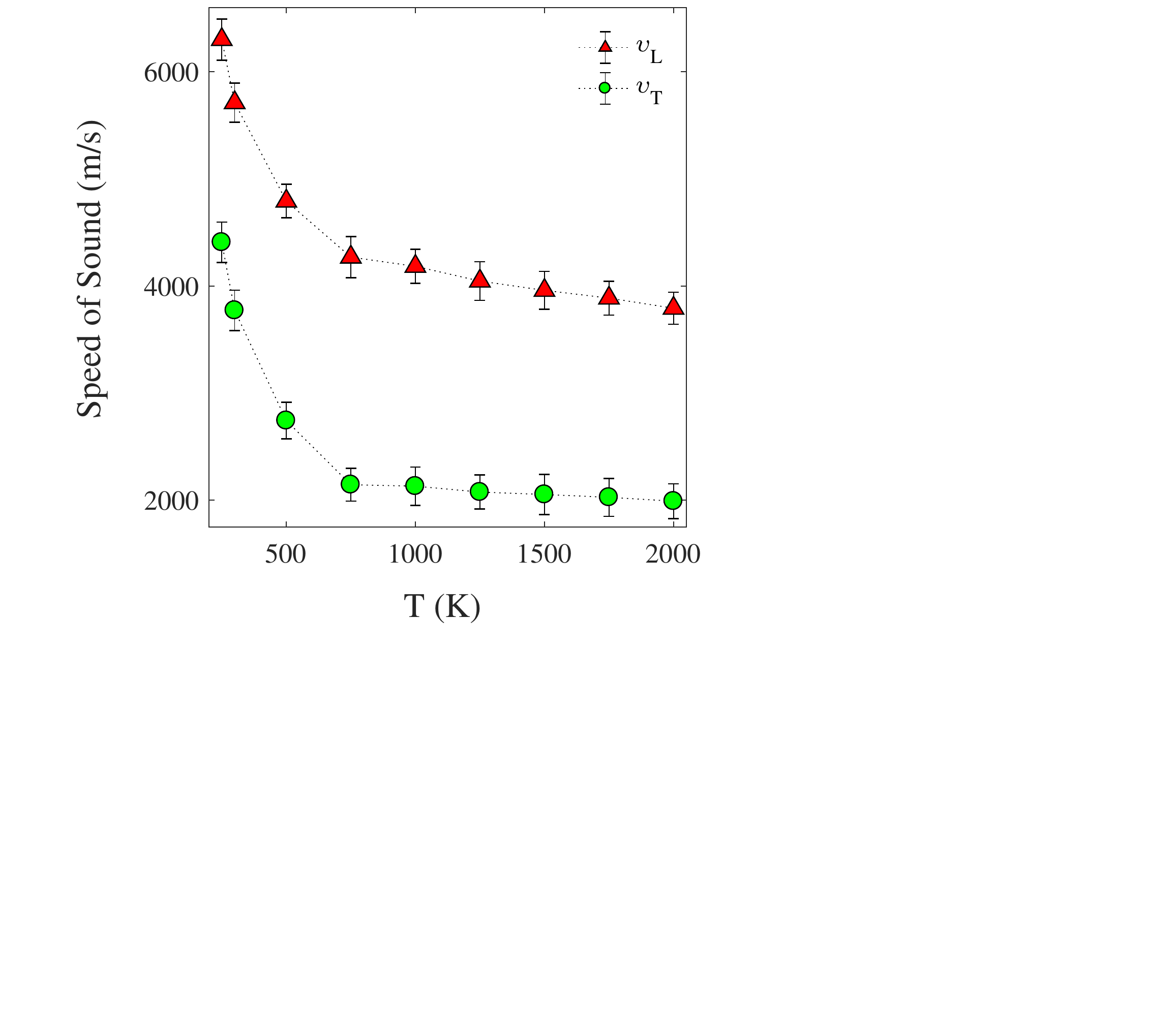}
	\caption{Temperature dependences of the longitudinal and transverse sound velocities for  Zr$_{50}$Cu$_{40}$Ag$_{10}$ metallic alloy. }
	\label{fig_3}
\end{figure}

The elastic characteristics and the velocities of the longitudinal and transverse ultrasonic waves for an isotropic medium are related by the expressions~\cite{tempbib14,tempbib15,tempbib16}
\begin{equation}
\label{eq:1.5}
     \vartheta_{L} =\sqrt{\frac{B + 4/3 G}{\rho}} , \; \; \; \;  \vartheta_{T} =\sqrt{\frac{G}{\rho}} \nonumber
\end{equation}
where $\rho$ is the mass density of the system. Fig. \ref{fig_3} depicts the obtained temperature dependences of the speed of sound for Zr$_{50}$Cu$_{40}$Ag$_{10}$ metallic alloy. As seen from the figure, that the temperature dependences of the longitudinal and transverse sound velocities for the equilibrium melt phase are monotonic. At the same time, near the liquid-glass transition, the changes in the temperature dependencies of both the rates $\vartheta_{L}$ and $\vartheta_{T}$ are observed.

\section{Summary}

In the present work, using molecular dynamics simulations, the elastic properties and glass forming ability of the ternary metallic alloy are investigated. The temperature dependences of bulk modulus $B$, the shear modulus $G$, the Young's modulus $E$ and the Poisson's ratio $\sigma$, for the Zr$_{50}$Cu$_{40}$Ag$_{10}$ metallic alloy were obtained for the first time. It is shown that the liquid-glass transition is accompanied by a considerable increase in the shear modulus $G$ and the Young's modulus $E$ (by more than $50\%$). Temperature dependence of the Poisson's ratio for the Zr$_{50}$Cu$_{40}$Ag$_{10}$-system exhibits a sharp fall from typical values for metals of approximately $0.32-0.33$ to low values (close to zero), which are characteristic for brittle bulk metallic glasses. It was also found that the temperature dependence of the sound velocities undergo significant changes near the liquid-glass transition. The calculated value of fragility index $m\approx64$ for the Zr$_{50}$Cu$_{40}$Ag$_{10}$-system is in a good agreement with the experimental data for the Zr-Cu-based metallic glasses.

\section*{Acknowledgements}
\vspace{\baselineskip} This work is supported by the Russian Science Foundation (project No. 19-12-00022). The molecular dynamic simulations were performed by using the computational cluster of Kazan Federal University and the computational facilities of Joint Supercomputer Center of RAS.

\end{document}